# Multi-Hop Cluster based IEEE 802.11p and LTE Hybrid Architecture for VANET Safety Message Dissemination

Seyhan Ucar*, Sinem Coleri Ergen† & Oznur Ozkasap*
*Department of Computer Engineering
†Department of Electrical & Electronics Engineering
Koc University, Turkey
sucar@ku.edu.tr, sergen@ku.edu.tr, oozkasap@ku.edu.tr

*Abstract*—Several Vehicular Ad hoc Network (VANET) studies have focused on the communication methods based on IEEE 802.11p, which forms the standard for Wireless Access for Vehicular Environments (WAVE). In the networks employing IEEE 802.11p only, the broadcast storm and disconnected network problems at high and low vehicle densities respectively degrade the delay and delivery ratio of safety message dissemination. Recently, as an alternative to the IEEE 802.11p based VANET, the usage of cellular technologies has been investigated due to their low latency and wide range communication. However, a pure cellular based VANET communication is not feasible due to the high cost of communication between the vehicles and the base stations, and high number of hand-off occurrences at the base station considering the high mobility of the vehicles. This paper proposes a hybrid architecture, namely VMaSC-LTE, combining IEEE 802.11p based multi-hop clustering and the fourth generation cellular system, Long Term Evolution (LTE), with the goal of achieving high data packet delivery ratio and low delay while keeping the usage of the cellular architecture at minimum level. In VMaSC-LTE, vehicles are clustered based on a novel approach named VMaSC: Vehicular Multi-hop algorithm for Stable Clustering. The features of VMaSC are cluster head selection using the relative mobility metric calculated as the average relative speed with respect to the neighboring vehicles, cluster connection with minimum overhead by introducing direct connection to the neighbor that is already a head or member of a cluster instead of connecting to the cluster head in multiple hops, disseminating cluster member information within periodic hello packets, reactive clustering to maintain cluster structure without excessive consumption of network resources, and efficient size and hop limited cluster merging mechanism based on the exchange of the cluster information among the cluster heads. These features decrease the number of cluster heads while increasing their stability therefore minimize the usage of the cellular architecture. From the clustered topology, elected cluster heads operate as dual-interface nodes with the functionality of IEEE 802.11p and LTE interface to link VANET to LTE network. Using various key metrics of interest including data packet delivery ratio, delay, control overhead and clustering stability, we demonstrate superior performance of the proposed architecture compared to both previously proposed hybrid architectures and alternative routing mechanisms including flooding and cluster based routing via extensive simulations in ns-3 with the vehicle mobility input from the Simulation of Urban Mobility (SUMO). The proposed architecture also allows achieving higher required reliability of the application quantified by the data packet delivery ratio at the cost of higher LTE usage determined by the number of cluster heads in the network.

## I. INTRODUCTION

VANET is expected to significantly improve the safety of our transportation systems by providing timely and efficient data dissemination about events like accidents, road conditions and traffic jams beyond the driver's knowledge [2]. Driver behavior, constraints on mobility and high speeds create unique characteristics such as rapid but somewhat predictable topology changes, uneven network density and frequent fragmentation for VANET. Meeting the strict delay and packet delivery requirements of safety applications in such a dynamic network determines the feasibility of the deployment of such applications. Table I shows the specifications of various VANET safety applications extracted from [3], [4]: update rate refers to the packet generation rate of the nodes; maximum dissemination distance is defined as the distance within which the safety message needs to be disseminated; maximum delay is the maximum tolerable delay for the safety message dissemination. The packet delivery ratio of the safety application, which is measured as the ratio of the nodes that successfully receive packets within the maximum dissemination distance, on the other hand mostly ranges from 90% to 100% depending on the application type and network scenario although it is not provided explicitly in the safety application specifications.

Up to now, several VANET studies have focused on the communication methods based on IEEE 802.11p, which forms the standard for WAVE. IEEE 802.11p provides data rate ranging from 6 Mbps to 27 Mbps at short radio range, around 300 m. Disseminating safety information over a large area requires an intelligent multi-hop broadcast mechanism handling two major problems: broadcast storm [5] and disconnected network [6]. The broadcast storm problem occurs at high vehicle traffic density where the packet delay and number of collisions at the medium access control layer increase dramatically as the number of vehicles attempting to transmit simultaneously increases. Probabilistic flooding [5] and clustering [7]–[19] are commonly used to address the broadcast storm problem. On the other hand, the disconnected network problem occurs at low vehicle traffic density where the number of nodes is not sufficient to disseminate the information to all the vehicles in a certain region. Store-carry-forward, where the vehicles in the opposite lane are used for message dissemination, is commonly utilized to address the disconnected network problem [6], [20]. The solutions addressing both broadcast storm and disconnected network problems however have been shown to provide network delays varying from a few seconds to several minutes and the percentage of the vehicles successfully receiving the packets going down to $60\%$ [21].

[1]This work has been conducted as part of Turk Telekom Research project under Grant Number 11315-07. Sinem Coleri Ergen also acknowledges support from Bilim Akademisi - The Science Academy, Turkey under the BAGEP program. A preliminary version of this work appeared in IEEE Wireless Communications and Networking Conference, Shanghai, China, Apr. 2013 [1]

Recently, as an alternative to the IEEE 802.11p based VANET, the usage of cellular technologies has been investigated. The key enabler of such usage is the standardization of the advanced content broadcast/multicast services by the Third Generation Partnership Project (3GPP), which provides efficient message dissemination to many users over a geographical area at fine granularity. The use of the third generation mobile cellular system, called Universal Mobile Communication System (UMTS), in the safety application of the vehicles has already been experimented in Project Cooperative Cars (CoCars) [22]. The traffic hazardous warning message has been shown to be disseminated in less than one second. The fourth generation cellular system, called Long Term Evolution (LTE), is an evolution of UMTS increasing the capacity and speed using a different radio interface together with core network improvements. The LTE specification provides down-link peak rates of 300 Mbps, up-link peak rates of 75 Mbps, transfer latency of less than 5 ms and transmission range up to 100 km in the radio access network. Despite the high rate coupled with wide-range communication, however, a pure LTE based architecture is not feasible for vehicular communication due to the high cost of LTE communication between the vehicles and the base stations, and high number of hand-off occurrences at the base station considering the high mobility of vehicles [23], [24]. Moreover, LTE architecture may overload due to heavy traffic broadcasted in dense areas where LTE is unable to fulfill the delivery requirement of safety application [25].

Hybrid architectures have been recently proposed to exploit both the low cost of IEEE 802.11p and the wide range low latency communication of the cellular technologies as summarized in Table II. Some of these works [26], [29], [34] focus on the usage of the hybrid architecture for more efficient clustering: [26] demonstrates the usage of the cellular communication signaling in the hybrid architecture; [29] exploits the usage of the centralized architecture of the cellular communication to reduce the clustering overhead; [34] proposes a new protocol based on the selection of a route with the longest lifetime to connect to the wired network for services such as driver information systems and Internet access. On the other hand, [31]–[33] propose cluster based hybrid architecture for message dissemination. In this hybrid architecture, the cluster members communicate with the cluster head by using IEEE 802.11p and the cluster heads communicate with the base station by using cellular technologies. The goal is to minimize the number of cluster heads communicating with the cellular network. Decreasing the number of clusters reduces the cost of using cellular infrastructure by lowering both the amount of communication with the base stations and the frequency of hand-off occurrences at the base station. Efficient clustering however should not only minimize the number of cluster heads but also maintain the stability of the cluster based topology with minimum overhead. None of the proposed hybrid architectures nevertheless perform any stability analysis. Moreover, [31] does not consider the delay performance of the message dissemination in the network. Although [32], [33] provide the delay performance of the hybrid architecture, they do not include the effect of multi-hop clustering on the number of cluster heads and the clustering stability. Furthermore, none of the previous hybrid architectures compare their performance to that of IEEE 802.11p based alternative routing mechanisms such as flooding and cluster based routing.

In the literature, VANET clustering has been performed with different purposes such as load balancing, quality of service support and information dissemination at high density vehicular networks [37]. Stable clustering with minimum number of cluster heads and minimum overhead requires efficient cluster joining, maintenance and merging mechanisms together with an efficient clustering metric considering the high mobility of vehicles. Clustering metrics used in the VANET literature include direction [7], [10]–[12]; packet delay variation [9]; location difference [8], [13], [15], [19]; speed difference [17]; combination of location and speed differences [14], [16], [18]. Although a metric combining the location and speed of the neighboring vehicles is a better measure of their link duration compared to a metric considering their speed only, all vehicles may not have localization capability. Calculating packet delay variation on the other hand requires very accurate synchronization among the vehicles with low level time stamping of the packets due to the random access protocol used by IEEE 802.11p. Besides, cluster joining in both one-hop and multi-hop VANET is directly to the cluster head. However, joining to the cluster through a cluster member and informing the cluster head later via periodic hello packets can decrease clustering connection time and overhead significantly. Such efficient mechanisms have been proposed in mobile ad hoc networks (MANET), which however usually assume stationarity of the nodes during clustering [38]. In addition, cluster maintenance is achieved through either periodic re-clustering [7]–[9], [11], [15], [16] where clustering procedure is executed periodically or reactive clustering [13], [14], [17] where clustering is triggered only when the cluster head has lost connection to all its members or cluster member cannot reach its cluster. Reactive clustering is more efficient since reclustering procedure is activated only when the cluster structure is destroyed without excessive periodic packet transmission overhead. Furthermore, previously proposed cluster merging mechanisms are activated either when the distance between two neighboring cluster heads is less than a certain threshold [11], [14], [17] or when the cluster heads remain connected for a time duration greater than a predetermined value [18], [19]. However, cluster merging can result in very large size merged clusters where cluster head becomes bottleneck due to the high number of packets of its cluster members and large number of hops that increases the delay of packet transmissions. To solve the cluster head bottleneck and large delay problems, cluster merging should limit both the size and number of hops in the resulting merged cluster. Also, previously proposed multi-hop clustering algorithms only focus on providing clustering stability through metrics such as cluster head duration, cluster member duration and cluster head change but do not analyze the performance of their proposed algorithm in message dissemination in terms of metrics such as packet delivery ratio and delay.

TABLE I: VANET Safety Application Requirements

| Service | Update Rate | Maximum Dissemination Distance | Maximum Delay |
|---|---|---|---|
| Safety Recall Notice | - | 400 m | 5 s |
| Vehicle Diagnostics and Maintenance | 10Hz | 500 m | 5 s |
| Wrong Way Driver Warning | 10Hz | 500 m | 1 s |
| Emergency Vehicle Signal Preemption | 1Hz | 1000 m | 1 s |
| Approaching Emergency Vehicle Warning | - | 1000 m | 1 s |

TABLE II: Related Work on Hybrid Architectures in VANETs

| Reference | Cellular Technology | Clustering | Radius | Application | Mobility Traces | Performance Criteria |
|---|---|---|---|---|---|---|
| [26] | UMTS | No | - | Control info dissemination | TraNS [27] and SUMO [28] | Packet delivery ratio, Packet drop, Delay |
| [29] | LTE | Yes | One-Hop | Cluster management | VanetMobiSim [30] | Packet loss, Overhead, Goodput, Cluster lifetime |
| [31] | UMTS | Yes | One-Hop | Message dissemination | - | Packet delivery ratio, Overhead, Packet loss |
| [32] | UMTS | Yes | One-Hop | Message dissemination | - | Packet delivery ratio, Overhead, Packet loss, Delay |
| [33] | LTE | Yes | One-Hop | Message dissemination | - | Packet delivery ratio, Packet error rate, Delay |
| [34] | UMTS | No | - | Connection to cellular backbone | MOVE [35] | Packet delivery ratio, Delay, Duplication ratio |

TABLE III: Related Work on VANET Clustering

| Reference | Hop Distance | Clustering Metric | Cluster Joining | Cluster Update | Cluster Merge | Mobility Traces | Performance Criteria |
|---|---|---|---|---|---|---|---|
| [7] | One-Hop | Direction | Directly to CH | Periodic | - | - | Packet delivery ratio, Clustering overhead |
| [8] | One-Hop | Location Difference | Directly to CH | Periodic | - | - | Packet delivery ratio |
| [10] | One-Hop | Direction | Directly to CH | - | - | Car following model | Packet reduction in transmission, Collision rate |
| [11] | One-Hop | Direction | Directly to CH | Periodic | Location Threshold | Car following model | Head duration, Head relative speed |
| [12] | One-Hop | Direction | Directly to CH | - | - | Car following model | Head change |
| [13] | One-Hop | Location Difference | Directly to CH | Reactive | - | Reference region group mobility model [36] | Head duration, Head change |
| [14] | One-Hop | Location Difference, Speed Difference | Directly to CH | Reactive | Location Threshold | Car following model | Cluster lifetime, Number of clusters |
| [15] | One-Hop | Location Difference | Directly to CH | Periodic | - | - | Head duration, Member duration, Connectivity |
| [17] | One-Hop | Speed Difference | Directly to CH | Reactive | Location Threshold | - | Head Change, Throughput |
| [18] | One-Hop | Location Difference, Speed Difference | Directly to CH | - | Merge Timer | MOVE [35] | Head duration, Member Duration, Head change,, Number of clusters |
| [19] | One-Hop | Location Difference | Directly to CH | - | Merge Timer | SUMO [28] | Cluster Life Time |
| [9] | Multi-Hop | Delay Variation | Directly to CH | Periodic | - | Free way and Manhattan model | Head duration, Member duration, Head change |
| [16] | Multi-Hop | Location Difference, Speed Difference | Directly to CH | Periodic | - | VanetMobiSim [30] | Number of clusters, Clustering overhead |

In this paper, we propose a hybrid architecture, namely VMaSC-LTE, combining IEEE 802.11p based multi-hop clustering and LTE with the goal of achieving high data packet delivery ratio and low delay while keeping the usage of the cellular infrastructure at minimum level via minimizing the number of cluster heads and maximizing the clustering stability. The original contributions of the paper are listed as follows:

- We propose a multi-hop cluster based IEEE 802.11p-LTE hybrid architecture for the first time in the literature. The features of the multi-hop clustering algorithm used in this hybrid architecture, called VMaSC, are cluster head selection using the relative mobility metric calculated as the average relative speed with respect to the neighboring vehicles, cluster connection with minimum overhead by introducing direct connection to the neighbor that is already a head or member of a cluster instead of connecting to the cluster head in multiple hops, disseminating cluster member information within periodic hello packets, reactive clustering to maintain cluster structure without excessive consumption of network resources, and efficient size and hop limited cluster merging mechanism based on the exchange of the cluster information among the cluster heads. Combining all of these features in a multi-hop cluster based hybrid architecture, using minimum overhead cluster connection, and size and hop limited cluster merging mechanism are unique characteristics of VMaSC.
- We perform an extensive analysis of the performance of the multi-hop cluster based IEEE 802.11p-LTE hybrid architecture over a wide range of performance metrics including data packet delivery ratio, delay, control overhead and clustering stability in comparison to both previously proposed hybrid architectures and alternative routing mechanisms including flooding and cluster based routing over a large scale highway using a realistic vehicle mobility model for the first time in the literature.
- We illustrate the trade-off between the reliability of the application measured by the data packet delivery ratio and the cost of the LTE usage determined by the number of cluster heads in the network for the first time in the literature.

The rest of the paper is organized as follows. Section II describes the system model. Section III presents the proposed multi-hop clustering scheme. Section IV delineates the data forwarding approach in the IEEE 802.11p-LTE hybrid architecture. The comparison of the proposed hybrid architecture to the previously proposed hybrid architectures and alternative routing mechanisms is given in Section V. Finally, concluding remarks and future work are given in Section VI.

## II. SYSTEM MODEL

The envisioned IEEE 802.11p-LTE hybrid architecture is shown in Fig. 1. The vehicles form a multi-hop clustered topology in each direction of the road. The vehicles within the transmission range of a cluster head (CH), which is denoted by $R$ and shown by dotted line around CH in the figure, become cluster member (CM) and directly communicate with their corresponding CH. The vehicles that are multi-hop away from the CH become multi-hop CMs and transfer data packets to the CM they are connected to in order to reach their corresponding CH.

The vehicle information base ($VIB$) of a vehicle consists of a repository storing the information of the vehicle and its neighboring vehicles within a predetermined maximum number of hops, denoted by $MAX\_HOP$. $VIB$ is used in determining the members and heads of the clusters in the network.

The vehicles possess two communication interfaces: IEEE 802.11p and LTE. CMs can only communicate with the members of the cluster they belong to via IEEE 802.11p whereas CH communicates with both CMs via IEEE 802.11p and eNodeB via LTE.

The LTE infrastructure is responsible for disseminating the generated data within VANET inside a geographical region. LTE part of the system consists of a radio access network (RAN) where each cell is managed by an eNodeB and the Evolved Packet Core (EPC) which consists of server gateway (SGW) and Packet Data Network Gateways (PGW) [39]. eNodeB is a complex base station that handles radio communications with multiple devices in the cell and carries out radio resource management and handover decisions. SGW provides the functionality of routing and forwarding data packets to neighboring eNodeBs whereas PGW is responsible for setting the transfer paths of vehicle data packets, quality of service control and authentication. eNodeBs are connected to EPC over a wired network. EPC has global information of the location of eNodeBs. When a CH sends the data packet to the eNodeB it is connected to over a radio network, the packet is sent to the EPC over the wired network. The EPC then determines all the eNodeBs that cover an area within the safety dissemination region of the data packet and sends the packet to them. When an eNodeB receives a data packet for dissemination, the packet is multicast to all the CHs that are within the coverage of eNodeB.

The objective of the proposed hybrid architecture is to efficiently forward data packets over a certain geographical region with small delay and high percentage of vehicles successfully receiving packets while minimizing the number of cluster heads and maximizing the clustering stability to minimize the overhead on the vehicles and eNodeB.

## III. VEHICULAR MULTI-HOP ALGORITHM FOR STABLE CLUSTERING (VMASC)

The features of the proposed multi-hop clustering algorithm VMaSC are as follows:

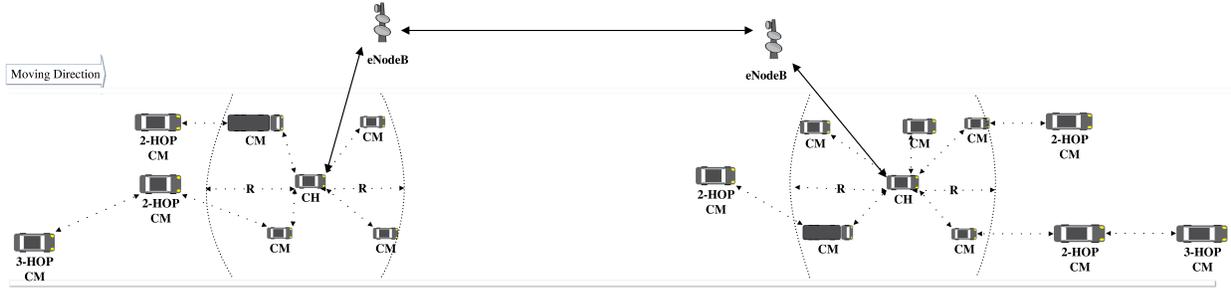

Fig. 1: IEEE 802.11p-LTE Hybrid Architecture

1) It provides stable cluster head selection by the use of the relative mobility metric calculated as the average relative speed with respect to the neighboring vehicles in multi-hop clustered vehicular network.
2) It provides cluster connection with minimum overhead by introducing direct connection to the neighbor that is already a head or member of a cluster, instead of connecting to the cluster head in multiple hops and disseminating cluster member information within periodic hello packets.
3) It provides reactive clustering to maintain cluster structure without excessive packet transmission overhead.
4) It provides minimum inter-cluster interference by minimizing the overlap of clusters in space through prioritizing the connections to existing clusters and introducing efficient size and hop aware cluster merging mechanisms based on the exchange of the cluster information among the cluster heads.

Fig. 1 shows an example multi-hop clustered network topology. Next, we describe the states of the vehicles, $VIB$ generation and update, cluster state transitions, cluster formation, cluster merging and inter-cluster interference. The notation used is presented in Table IV.

TABLE IV: Notation

| Notation | Description |
|---|---|
| $IN$ | Initial State |
| $SE$ | State Election |
| $CH$ | Cluster Head |
| $ISO-CH$ | Isolated Cluster Head |
| $CM$ | Cluster Member |
| $VIB$ | Vehicle Information Base |
| $VIB\_TIMER$ | VIB Timer |
| $IN\_TIMER$ | Initial State Timer |
| $SE\_TIMER$ | State Election State Timer |
| $CH\_TIMER$ | Cluster Head State Timer |
| $CM\_TIMER$ | Cluster Member State Timer |
| $JOIN\_TIMER$ | Join Response Packet Timer |
| $MERGE\_TIMER$ | Merge Timer |
| $V_{state}$ | Vehicle's Current State |
| $AVGREL\_SPEED_i$ | Average Relative Speed of Vehicle $i$ |
| $MEMBER_{CH}$ | CH's Connected Member Counter |
| $MEMBER_{CM}$ | CM's Connected Member Counter |
| $MAXMEMBER\_CH$ | Max. Member CH can serve |
| $MAXMEMBER\_CM$ | Max. Member CM can serve |
| $MAX\_HOP$ | Max. Hop Between CH and CM |
| $HOP_{CM}$ | Number of hops between CM and CH |
| $CH\_ADV$ | CH's Advertisement Packet |
| $JOIN\_REQ$ | Join Request Packet |
| $JOIN\_RESP$ | Join Response Packet |
| $CH\_ADV$ | Cluster Head Advertisement Message |
| $HELLO\_PACKET$ | Periodic Hello Packet |
| $DATA\_PACKET$ | Data Packet |
| $CLUSTER\_INFO$ | Cluster Information Packet |
| $TRY\_CONNECT_i$ | Try Connect Flag for Vehicle $i$ |
| $MERGE\_REQ$ | CH's Merge Request |
| $MERGE\_RESP$ | CH's Merge Response |
| $ID_{DATA}$ | Data Packet Generator Identifier |
| $SEQ_{DATA}$ | Data Packet Sequence Number |
| $PARENT_i$ | Vehicle through which vehicle $i$ is connected to the cluster |
| $CHILDREN_i$ | Vehicles that use vehicle $i$ to connect to the CH |

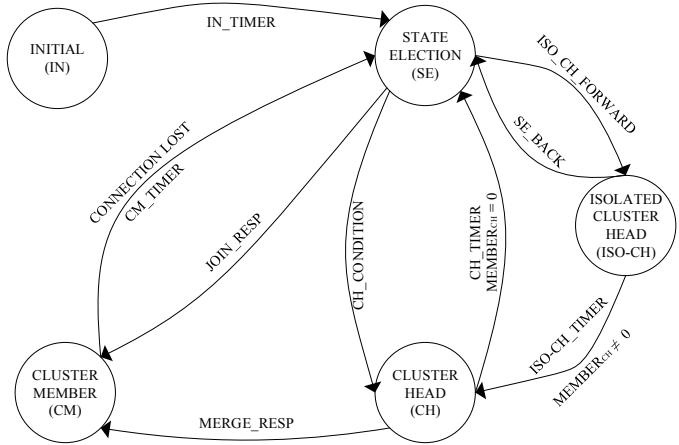

Fig. 2: VMaSC State Transition Diagram

### A. States of Vehicles

Each vehicle operates under one of the following five states at any given time:

- INITIAL ($IN$) is the starting state of the vehicle.
- STATE ELECTION ($SE$) is the state of the vehicle in which the vehicle makes a decision about the next state based on the information in $VIB$.
- CLUSTER HEAD ($CH$) is the state of the vehicle in which the vehicle is declared to be cluster head.
- ISOLATED CLUSTER HEAD ($ISO-CH$) is the state

the vehicle makes a transition to when it cannot connect to any existing cluster and there is no potential neighboring vehicle that can connect to it.
- CLUSTER MEMBER ($CM$) is the state of the vehicle in which the vehicle is attached to an existing cluster.

### B. VIB Generation and Update

$VIB$ at each node includes the information of the vehicle itself and its neighboring vehicles within $MAX\_HOP$ hops. The vehicle information includes its direction, its velocity, its current clustering state, the number of hops to the cluster head if it is a cluster member, the ID of the vehicle through which the node is connected to the cluster, the ID of the vehicles that use the node to connect to the cluster head, its clustering metric, and the source ID and sequence number of the data packets that are recently generated.

$VIB$ is updated upon any change in the vehicle's own information or reception of a periodic $HELLO\_PACKET$ from any of the neighbors within $MAX\_HOP$ hops. $HELLO\_PACKET$ includes the vehicle information for its direction, its velocity, its current clustering state, the number of hops to the cluster head if it is a cluster member and the ID of the vehicle through which the node is connected to the cluster. $HELLO\_PACKET$ is re-transmitted to the neighbors within $MAX\_HOP$ hops. The entries of $VIB$ are deleted if not updated for a certain time denoted by $VIB\_TIMER$.

The clustering metric denoted by $AVGREL\_SPEED_i$ for vehicle $i$ is calculated as

$$AVGREL\_SPEED_i = \frac{\sum_{j=1}^{N(i)} |S_i - S_{i_j}|}{N(i)} \quad (1)$$

where $N(i)$ is the number of same direction neighbors within $MAX\_HOP$ hops for vehicle $i$, $i_j$ is the ID of the $j$-th neighboring node of vehicle $i$, $S_i$ is the speed of vehicle $i$. The lower the average relative speed, the less mobile the vehicle compared to its neighbors. Therefore, the vehicle with the lowest average relative speed is elected as the cluster head.

### C. Cluster State Transitions

Fig. 2 illustrates the possible state transitions of a vehicle. The vehicle starts in state $IN$ and stays in this state for a duration denoted by $IN\_TIMER$. The periodic exchange of $HELLO\_PACKET$ in this state helps the vehicle build its own $VIB$. The vehicle then transitions to state $SE$ in which it makes the decision about its next state as described in Algorithm 1 in Section III-D.

A vehicle transitions from state $SE$ to state $CM$ if it receives a $JOIN\_RESP$ from a CH or CM. Receiving a $JOIN\_RESP$ shows the success of the join request. A vehicle transitions from state $SE$ to $CH$ if $CH\_CONDITION$ is satisfied. $CH\_CONDITION$ refers to the condition where the vehicle cannot connect to any neighboring CH or CM, there is at least one neighboring vehicle in state $SE$ and the vehicle has minimum average relative speed among all the neighboring vehicles in state $SE$. The vehicle in state $SE$ transitions to state $ISO-CH$ if $ISO\_CH\_FORWARD$ condition is met. $ISO\_CH\_FORWARD$ refers to the condition where the vehicle cannot connect to any neighboring CH or CM and there is no neighboring vehicle in state $SE$. The vehicle in state $ISO-CH$ behaves identical to state $CH$. A separate $ISO-CH$ state is included to differentiate the CHs without any cluster members since the condition to switch from $CH$ to $SE$ is the lack of a cluster member connected to CH. If the number of the members of the $CH$ denoted by $MEMBER\_CH$ is zero for the duration $CH\_TIMER$, the CH changes its state to $SE$ in order to decrease the number of clusters in the network by connecting to another cluster. The vehicle in state $ISO-CH$ transitions to state $CH$ if a cluster member is connected after $ISO\_CH\_TIMER$ and to state $SE$ when the node is not isolated any more, meeting the $SE\_BACK$ condition. $SE\_BACK$ refers to the condition of discovering a node in either $CM$ or $CH$ state that does not exist in the $VIB$. If none of the transitions can be made in state $SE$, then the vehicle stays in $SE$ for a duration denoted by $SE\_TIMER$, then rerun Algorithm 1.

A vehicle changes state from $CH$ to $CM$ if it receives a $MERGE\_RESP$ from another CH demonstrating the success of the cluster merging as explained in detail in Section III-E. The vehicle transitions from state $CM$ to $SE$ if it has lost connection to the neighboring node through which it is connected to the cluster named $PARENT$. If CM vehicle does not receive any packet from its $PARENT$ for the duration $CM\_TIMER$, it assumes to have lost the connection.

### D. Cluster Formation

As shown in detail in Algorithm 1, a vehicle in $SE$ state first tries to connect to the existing clusters in order to minimize the number of cluster heads. The vehicle gives priority to the neighboring CHs over the neighboring CMs for connection to decrease the delay of the data packets transmitted to the CH over smaller number of hops.

The vehicle first scans the neighboring CHs in the order of increasing average relative mobility. If the number of members of the CH is less than the maximum number of members allowed and the vehicle has not tried connecting to that CH before with $TRY\_CONNECT$ set to false, it sends a $JOIN\_REQ$ packet (Lines $1-4$). If it receives $JOIN\_RESP$ from the corresponding CH within a given amount of time denoted by $JOIN\_TIMER$, the vehicle transitions to state $CM$ and exits the $SE$ algorithm (Lines $5-7$). Otherwise, the vehicle sets the $TRY\_CONNECT_{CH}$ flag to $true$ in order not to try connecting to that CH again (Line 9). $TRY\_CONNECT$ flags of vehicles are initially set to false.

If none of the neighboring vehicles is a CH or the vehicle cannot connect to any of the neighboring CHs, then the vehicle tries to connect to a CH in multiple hops through a CM (Lines $10-19$). The order in which the CMs are scanned is determined based on the average relative mobility. Similar to the connection to CH, if the number of members of the CM is less than the maximum number of members allowed and the vehicle has not tried connecting to that CM

**Algorithm 1:** State Election (SE) Algorithm

**1 forall the** $CH \in VIB$ **do**
**2**    **if** $TRY\_CONNECT_{CH} == false$ **then**
**3**       **if** $MEMBER_{CH} < MAXMEMBER\_CH$ **then**
**4**          Send $JOIN\_REQ$;
**5**          **if** $JOIN\_RESP$ received **then**
**6**             $V_{state} = CM$;
**7**             Exit;
**8**          **else**
**9**             $TRY\_CONNECT_{CH} = true$;

**10 forall the** $CM \in VIB$ **do**
**11**    **if** $TRY\_CONNECT_{CM} == false$ **then**
**12**       **if** $MEMBER_{CM} ¡ MAXMEMBER\_CM$ **then**
**13**          **if** $HOP_{CM} < MAX\_HOP$ **then**
**14**             Send $JOIN\_REQ$;
**15**             **if** $JOIN\_RESP$ received **then**
**16**                 $V_{state} = CM$;
**17**                 Exit;
**18**             **else**
**19**                 $TRY\_CONNECT_{CM} = true$;

**20 if** *Not exists* $SE \in VIB$ **then**
**21**    $V_{state} = ISO - CH$;
**22**    Exit;
**23 else if** $AVGREL\_SPEED_{curr} = min_{SE \in VIB}(AVGREL\_SPEED_{SE})$ **then**
**24**    $V_{state} = CH$;
**25**    Broadcast $CH\_ADV$;
**26**    Exit;

before with $TRY\_CONNECT$ set to false (Lines $10-12$), and the vehicle is within $MAX\_HOP$ hops away from the corresponding CH (Line 13), the vehicle sends $JOIN\_REQ$ packet to this CM (Line 14). Depending on the reception of the $JOIN\_RESP$, the vehicle then either transfers to state $CM$ or sets the try connect flag of the $CM$ to $true$ (Lines $15-19$). If the vehicle receives multiple $JOIN\_RESP$s then it prefers $CH$ over $CM$ and the vehicle with the smallest average relative mobility among multiple CHs and CMs.

If the vehicle cannot connect to any CH or CM, the vehicle checks the neighboring vehicles in $SE$ state. If there is no such vehicle, it transitions to state $ISO - CH$ (Lines $20-22$). If there are vehicles in $SE$ state in its $VIB$ and the vehicle has the smallest average relative speed, it makes a transition to $CH$ state and broadcasts $CH\_ADV$ packet (Lines $23-26$). Otherwise, the vehicle stays in state $SE$ for $SE\_TIMER$ duration and reruns Algorithm 1. This cluster formation technique constructs hierarchical organization between vehicles where an efficient aggregation algorithm can be implemented [40] and applied by vehicles before forwarding data packets to parent vehicle.

*E. Cluster Merging*

Since the vehicles do not send the $JOIN\_REQ$ messages to the CH in multiple hops, the CH learns about the vehicles within its cluster via the $HELLO\_PACKET$. The CH keeps the information about its cluster including the ID and $PARENT$ node of its cluster members and its cluster direction within a data structure named $CLUSTER\_INFO$.

When two CHs become neighbors, they first check whether they stay neighbors for a certain time period denoted by $MERGE\_TIMER$. The value of $MERGE\_TIMER$ should be chosen to balance the trade-off between cluster stability and number of clusters in the network: As $MERGE\_TIMER$ increases, the cluster stability increases at the cost of increase in the number of clusters. If the CHs stay neighbors for $MERGE\_TIMER$, they share their $CLUSTER\_INFO$ and their average relative speed with each other. Both CHs then check the feasibility of the merged cluster formed when the CH with higher average relative speed gives up its CH role and connects to the CH with lower average relative speed. A feasible merged cluster requires that both clusters have the same cluster direction, the number of members of the CH and CMs in the merged cluster be less than $MAXMEMBER\_CH$ and $MAXMEMBER\_CM$ respectively, and the number of hops in the merged cluster be less than $MAX\_HOP$. Limiting the maximum number of vehicles and the number of hops in the merged cluster eliminates cluster head bottleneck and longer hierarchical routes respectively. If the merged cluster is determined to be feasible then the CH with higher average relative speed sends $MERGE\_REQ$ to the less mobile CH. If this CH receives $MERGE\_RESP$, it gives up its CH role and informs its cluster members about the merge operation. Otherwise, the CHs continue to function as cluster heads. If the vehicle receives multiple $MERGE\_RESP$s then it prefers the CH with the smallest average relative mobility.

*F. Inter-cluster Interference*

Inter-cluster interference occurs when the clusters overlap in space. Inter-cluster interference leads to higher medium contention and inefficient flooding. VMaSC minimizes overlapping clusters via two methods. 1) The vehicles in $SE$ state try to join to an existing cluster first before declaring themselves as $CH$ or $ISO-CH$. 2) The CHs that are within the transmission range of each other merge their clusters if the resulting merged cluster is considered feasible. Moreover, the data packets of the CMs are unicast to their $PARENT$ to decrease the medium contention and increase the efficiency of the flooding. Furthermore, the packets of each cluster are only broadcast within that cluster identified with a unique ID avoiding unnecessary flooding among multiple clusters.

*G. Theoretical Analysis of VMaSC Clustering*

In this section, we provide the theoretical analysis of the relative speed metric used in the VMaSC clustering.

Let us assume that two neighboring vehicles 1 and 2 have average speed $v_1$ and $v_2$, and average acceleration $a_1$ and $a_2$, respectively. Assume that these vehicles move on a one dimensional road and they communicate with each other only if they are within the transmission range of each other called $r_t$. Let the location of the vehicles 1 and 2 on the road in the moving direction be $l_1$ and $l_2$ with difference denoted by $r_{12}$ equal to $l_1 - l_2$. $r_{12}$ is a random variable that takes values within $[-r_t, r_t]$ interval. The inter-vehicle distance has been shown to have exponential distribution at low vehicle density and log-normal distribution at high vehicle density [6]. In this case, it should also be conditioned on the fact that its value is in the $[-r_t, r_t]$ range. We represent the distribution of $r_{12}$ by $P(r_{12})$ without making any assumption on its distribution except that it is limited to $[-r_t, r_t]$ range and symmetric around 0. Since the vehicles exchange their speed information with each other, we assume that $v_1$ and $v_2$ are predetermined. The average acceleration of the vehicles $a_1$ and $a_2$ however are assumed to be random variables. Most theoretical analyses related to clustering in the literature assume no vehicle acceleration, i.e. $a_1 = a_2 = 0$ [41]–[43]. The Freeway mobility model on the other hand assumes $a_1$ and $a_2$ are independent random variables with uniform distribution in the interval $[-a, a]$, where $a$ is determined by the maximum acceleration and deceleration of the vehicles, while also enforcing the minimum and maximum speed values for the vehicles and the minimum safety distance between any two vehicles, generating a possibly non-zero correlation on the values of the accelerations $a_1$ and $a_2$; whereas Reference Point Group Mobility Model determines the speed of each vehicle by randomly deviating from the speed of a vehicle in their group as a function of a predetermined speed and angle deviation ratio [44], [45]. To encompass all these different mobility models while preserving the tractability of the analysis, we assume that the distribution of the difference between $a_1$ and $a_2$, $\delta_{12}$, denoted by $f(\delta_{12})$ is symmetric around 0 and a decreasing function of $\delta_{12}$ for $\delta_{12} > 0$.

Let us first condition on the values of $a_1$ and $a_2$. At any time $T$, if $r_{12} + (v_1 - v_2)T + (a_1 - a_2)T^2/2$ is less than $r_t$ in magnitude, the two vehicles can communicate with each other. In order to have a more stable link among the cluster members, at any given time $T$, we need to maximize the probability that the vehicles can communicate with each other given by

$$P(-r_t < r_{12} + (v_1 - v_2)T + (a_1 - a_2)T^2/2 < r_t)$$
$$= \int_{-r_t}^{r_t} P(-r_t - r_{12} < (v_1 - v_2)T + (a_1 - a_2)T^2/2$$
$$< r_t - r_{12} | r_{12} = \tau) P(\tau) d\tau \quad (2)$$

where $P(-r_t - r_{12} < (v_1 - v_2)T + (a_1 - a_2)T^2/2 < r_t - r_{12} | r_{12} = \tau)$ takes value 1 or 0 depending on the values of $\tau$, $v_1 - v_2$, $a_1 - a_2$, $r_t$ and $T$ parameters. By using the symmetry of the distribution $P(r_{12})$ around 0, this probability can be simplified as

$$\int_{-r_t}^{r_t - |(v_1 - v_2)T + (a_1 - a_2)T^2/2|} P(\tau) d\tau \quad (3)$$

If $a_1 = a_2$, then to maximize this connection probability, we need to minimize the relative speed of the vehicles given by $|v_1 - v_2|$. On the other hand, if $a_1 \neq a_2$, the probability that the two vehicles are in communication range with each other is given by

$$\int_{-\infty}^{\infty} P(-r_t < r_{12} + (v_1 - v_2)T + \delta_{12}T^2/2 < r_t) f(\delta_{12}) d\delta_{12}$$
$$= \int_{-\infty}^{\infty} \int_{-r_t}^{r_t - |(v_1 - v_2)T + \delta_{12}T^2/2|} P(\tau) f(\delta_{12}) d\tau d\delta_{12} \quad (4)$$

By using the symmetry of the distribution $f(\delta_{12})$ around 0, this probability can be simplified as

$$\int_0^{\infty} \left( \int_{-r_t}^{r_t - |(v_1 - v_2)T + \delta_{12}T^2/2|} P(\tau) d\tau + \right.$$
$$\left. \int_{-r_t}^{r_t - |(v_1 - v_2)T - \delta_{12}T^2/2|} P(\tau) d\tau \right) f(\delta_{12}) d\delta_{12} \quad (5)$$

By using the fact that both $|(v_1 - v_2)T + \delta_{12}T^2/2|$ and $|(v_1 - v_2)T - \delta_{12}T^2/2|$ are lower bounded by $||(v_1 - v_2)T| - |\delta_{12}T^2/2||$ and upper bounded by $|(v_1 - v_2)T| + |\delta_{12}T^2/2|$, the probability that the two vehicles are in communication range with each other is lower bounded by

$$2 \int_0^{\infty} \int_{-r_t}^{r_t - |(v_1 - v_2)T| - \delta_{12}T^2/2} P(\tau) f(\delta_{12}) d\tau d\delta_{12} \quad (6)$$

and upper bounded by

$$2 \int_0^{\infty} \int_{-r_t}^{r_t - ||(v_1 - v_2)T| - \delta_{12}T^2/2|} P(\tau) f(\delta_{12}) d\tau d\delta_{12} \quad (7)$$

To maximize both lower bound and upper bound of this probability, we need to again minimize the relative speed of the vehicles $|v_1 - v_2|$.

## IV. Data Dissemination in Hybrid Architecture

The goal of the proposed multi-hop cluster based IEEE 802.11p-LTE hybrid architecture is to disseminate the data generated in the network to all the vehicles within a geographical area with small delay and high data packet delivery ratio. LTE is used in this architecture to provide the connectivity of the nodes even when the IEEE 802.11p based network is disconnected within the dissemination distance, and improve the delay and reliability performance of the transmissions when the IEEE 802.11p based network has high node density leading to high medium access contention.

The data forwarding at a vehicle depends on its clustering state. If its clustering state is $SE$, the vehicle broadcasts the $DATA\_PACKET$ so that it reaches a member of a cluster in the network. If the clustering state of the vehicle generating or receiving a $DATA\_PACKET$ is $CM$ ($CH$), the vehicle runs Algorithm 2 (Algorithm 3). The data flow is as follows:

1) Unicast from CM to its CH (if the vehicle is a CM)
2) Broadcast from CH to all its members and to the eNodeB
3) Unicast from eNodeB to EPC
4) Multicast from EPC to the neighboring eNodeBs covering a part of the geographical area targeted for the dissemination of the $DATA\_PACKET$.

5) Multicast from eNodeBs to the CHs within their coverage.
6) Broadcast from the CHs to all its members.

As provided in Algorithm 2, if the CM generates or receives a $DATA\_PACKET$, then it checks its $VIB$ to determine whether the packet has already been received (Lines $2-3$). If the CM receives the packet for the first time, then it checks the source of the packet. If it is coming from its parent vehicle ($PARENT_{curr}$) in the cluster, then it multicasts the packet to all its children $CHILDREN_{curr}$ (Lines $4-5$). Otherwise, the packet is from either one of its children or another vehicle that is in $SE$ state. The vehicle then unicasts the packet to its parent vehicle for the dissemination of the packet to the corresponding CH (Lines $6-7$) and updates its $VIB$ for the packet (Line $8$).

Likewise, as provided in Algorithm 3, if the CH generates or receives a $DATA\_PACKET$ for the first time, it checks the source of the packet (Lines $1-3$). If the packet is coming from eNodeB, the CH broadcasts the $DATA\_PACKET$ to all the members of its cluster (Lines $4-5$). Otherwise, the packet comes from itself, one of its children or another vehicle in $SE$ state. In that case, it broadcasts the packet to the members of its cluster, creates an $LTE\_DATA\_PACKET$ containing the data of the received packet, forwards the $LTE\_DATA\_PACKET$ to the eNodeB (Lines $6-8$) and updates its $VIB$ for the packet (Line $9$).

Upon reception of the $LTE\_DATA\_PACKET$ from a CH, the eNodeB multicasts the packet to all the CHs within its coverage and forwards it to the EPC via a wired network. The EPC then determines all the eNodeBs that cover an area within the dissemination region of the corresponding packet. The eNodeBs that receive an $LTE\_DATA\_PACKET$ from EPC then multicast it to all the CHs under their coverage, which again disseminate the data to their CMs.

---

**Algorithm 2:** IEEE 802.11p-LTE CM Algorithm

1 On $DATA\_PACKET$ generation or receipt:
2 Extract $ID_{DATA}$ and $SEQ_{DATA}$;
3 **if** $(ID_{DATA}, SEQ_{DATA}) \notin VIB$ **then**
4    **if** $DATA\_PACKET$ is from $PARENT_{curr}$ **then**
5       Multicast $DATA\_PACKET$ to $CHILDREN_{curr}$;
6    **else**
7       Unicast $DATA\_PACKET$ to $PARENT_{curr}$;
8 Update $VIB$;

---

## V. Performance Evaluation

The simulations are performed in the Network Simulator ns3 (Release 3.17) [46] with the realistic mobility of the vehicles generated by SUMO [28]. SUMO, generated by the German Aerospace Center, is an open-source, space-continuous, discrete-time traffic simulator capable of modeling the behavior of individual drivers. The acceleration and over-taking decision of the vehicles are determined by using the distance to the leading vehicle, traveling speed, dimension of vehicles and profile of acceleration-deceleration.

---

**Algorithm 3:** IEEE 802.11p-LTE CH Algorithm

1 On $DATA\_PACKET$ generation or receipt:
2 Extract $ID_{DATA}$ and $SEQ_{DATA}$;
3 **if** $(ID_{DATA}, SEQ_{DATA}) \notin VIB$ **then**
4    **if** $DATA\_PACKET$ is from $eNodeB$ **then**
5       Broadcast $DATA\_PACKET$ into cluster;
6    **else**
7       Broadcast $DATA\_PACKET$ into cluster;
8       Put data in $LTE\_DATA\_PACKET$ and forward it to $eNodeB$;
9 Update $VIB$;

---

The goal of the simulations is to compare the performance of the proposed multi-hop cluster based IEEE 802.11p-LTE hybrid architecture to the previously proposed VANET multi-hop clustering algorithms NHop [9] and MDMAC [16], the hybrid architectures built with the usage of these clustering algorithms NHop and MDMAC, and flooding based message dissemination.

The road topology consists of a two-lane and two-way road of length 5 km. The vehicles are injected into the road according to a Poisson process with rate equal to 2 vehicles per second. The total simulation time is 355 s. The clustering process starts at 55 s when all the vehicles have entered the road. All the performance metrics are evaluated for the remaining 300 s. Two classes of vehicles with different maximum speed ranges are used in the simulation in order to create a realistic scenario with different types of vehicles on the road such as passenger cars, buses, trucks. The first vehicle class has maximum speed of 10 m/s whereas the maximum speed of the second vehicle class is considered as a variable ranging from 10 m/s to 35 m/s. Considering the injection of the vehicles into the road and their maximum speed constraint, the average number of the neighbors of the vehicles ranges from 10 to 18 at different times for different scenarios.

Tables V and VI list the simulation parameters of the VANET and LTE networks respectively. Maximum number of hops within a cluster is chosen in the range $[1, 3]$ since the number of hops above 3 reduces the clustering stability considerably due to the increase in the number of $HELLO\_PACKET$s disseminated within the maximum number of hops, increase in the number of retransmitted packets lost due to higher contention, and increase in the number of connections lost among the cluster members due to higher packet collisions. YANS channel model [47] used throughout the simulation is based on first deciding whether or not packet can be received at the beginning of the packet transmission considering the physical layer and Signal-to-Noise-Plus-Interference Ratio (SINR) level then determining the successful reception of the packet probabilistically at the end of the packet transmission by calculating the packet error rate as a function of the SINR level, modulation type, transmission rate and error correcting code.

We first compare the stability of the proposed clustering algorithm VMaSC to the previously proposed multi-hop VANET clustering algorithms. We then examine the delay and data packet delivery ratio performance of the proposed hybrid architecture compared to both previously proposed cluster based hybrid architectures and alternative mechanisms including flooding and pure VANET cluster based data forwarding.

TABLE V: ns-3 Simulation Parameters For VANET

| Parameters | Value |
| --- | --- |
| Simulation Time | 300 s |
| Maximum Velocity | 10 - 35 m/s |
| $MAX\_HOP$ | 1, 2, 3 |
| Number of Vehicles | 100 |
| Transmission Range | 200 m |
| MAC Layer | DCF, CSMA/CA |
| Channel Model | YANS [47] |
| $MAXMEMBER_{CH}$ | 5 |
| $MAXMEMBER_{CM}$ | 1 |
| $HELLO\_PACKET$ period | 200 ms |
| $HELLO\_PACKET$ size | 64 bytes |
| $DATA\_PACKET$ period | 1 s |
| $DATA\_PACKET$ size | 1024 bytes |
| $VIB\_TIMER$ | 1 s |
| $IN\_TIMER$ | 2 s |
| $SE\_TIMER$ | 2 s |
| $CH\_TIMER$ | 2 s |
| $CM\_TIMER$ | 2 s |
| $JOIN\_TIMER$ | 2 s |
| $MERGE\_TIMER$ | 2 s |
| $MAX\_HOP$ | 1, 2, 3 |

TABLE VI: ns-3 Simulation Parameters For LTE

| Parameters | Value |
| --- | --- |
| $eNodeB$ Scheduler Type | RrFfMacScheduler |
| $eNodeB$ Coverage | 7 km |
| Pathloss Model | Friis Propagation Model |

### A. VANET Clustering

In this section, VMaSC is compared to multi-hop clustering algorithms NHop [9] and MDMAC [16], the characteristics of which are summarized in Table III. The performance metrics used for comparison are cluster head duration, cluster member duration, cluster head change rate, clustering overhead and number of vehicles in $SE$ state.

*1) Cluster Head Duration:* Cluster head duration is defined as the time period from when a vehicle changes state to $CH$ to when a vehicle transitions from state $CH$ to state $SE$ or $CM$.

Fig. 3 shows the average cluster head duration of the clustering algorithms as a function of maximum vehicle velocities for different maximum number of hops whereas Fig.6 (a) shows the cumulative distribution function (cdf) of the cluster head duration of VMaSC for different maximum number of hops and different maximum vehicle velocities. The average cluster head duration of VMaSC is higher than that of the NHop and MDMAC under all conditions due to mainly efficient cluster maintenance mechanism based on reactive reclustering in VMaSC. Moreover, increasing the maximum number of hops allowed within each cluster increases the average cluster head duration and hence the clustering stability. The main reason is that the cluster head has a higher chance to find a member to serve as the number of hops increases. Besides, vehicles collect more information about surrounding vehicles at higher hops, which eventually contributes to better assignment of the roles $CH$ and $CM$ to the vehicles. Furthermore, the average cluster head duration in general decreases as the vehicle velocity increases since higher vehicle velocity results in higher dynamicity of the network topology. The cdf of the cluster head duration on the other hand shows that the variation around the average cluster head duration of VMaSC is small demonstrating its stability.

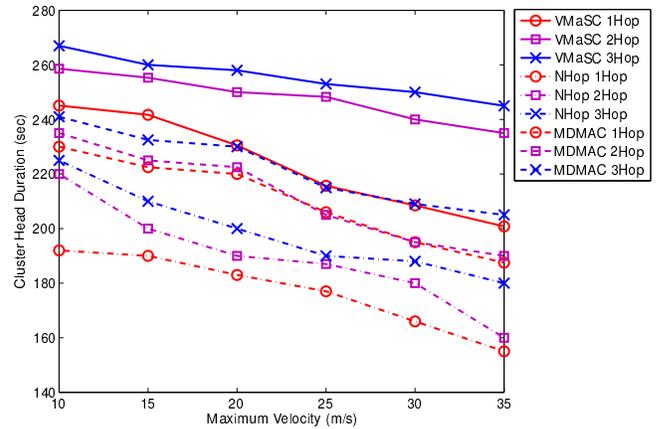

Fig. 3: Average cluster head duration of the clustering algorithms as a function of maximum vehicle velocities for different maximum number of hops.

*2) Cluster Member Duration:* Cluster member duration is defined as the time interval from joining an existing cluster as a member in $CM$ state to leaving the connected cluster by transitioning to $SE$ state.

Fig. 4 shows the average cluster member duration of the clustering algorithms as a function of maximum vehicle velocities for different maximum number of hops whereas Fig.6 (b) shows the cdf of the cluster member duration of VMaSC for different maximum number of hops and different maximum vehicle velocities. Similar to the cluster head duration, the average cluster member duration of VMaSC is higher than that of the NHop and MDMAC under all conditions due to again mainly the efficient cluster maintenance mechanism, and also low overhead and low delay cluster joining mechanism of VMaSC. The cluster member duration increases as the maximum number of hops increases since collecting more information about surrounding vehicles at higher number of hops enables the selection of better CH for connection. The cdf of the cluster member duration on the other hand shows that the variation around the average value of the cluster member duration of VMaSC is minimal similar to that of the cluster head duration.

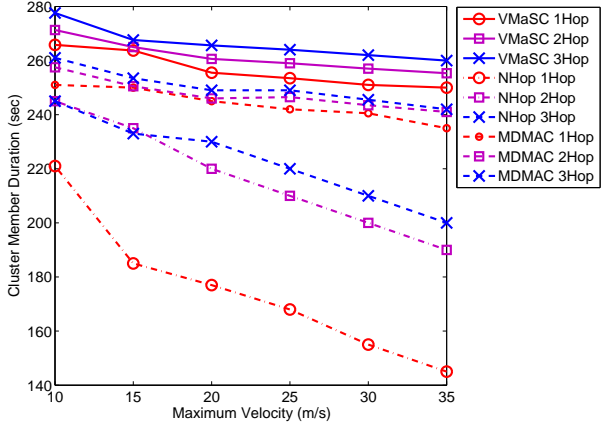

Fig. 4: Average cluster member duration of the clustering algorithms as a function of maximum vehicle velocities for different maximum number of hops.

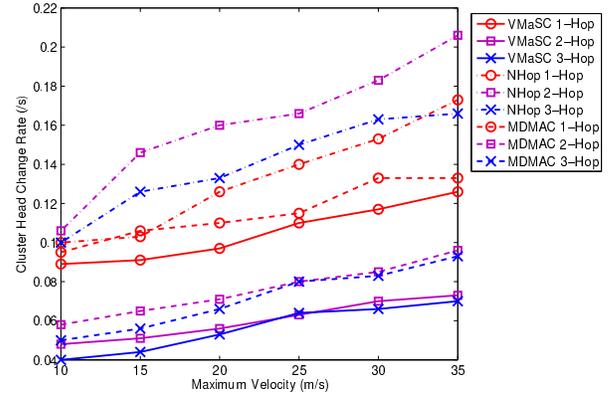

Fig. 5: Cluster head change rate of the clustering algorithms as a function of maximum vehicle velocities for different maximum number of hops.

*3) Cluster Head Change Rate:* Cluster head change rate is defined as the number of state transitions from $CH$ to another state per unit time.

Fig. 5 shows the cluster head change rate of the clustering algorithms as a function of maximum vehicle velocities for different maximum number of hops whereas Fig.6 (c) shows the cdf of the cluster head change rate of VMaSC for different maximum number of hops and different maximum vehicle velocities. The cluster head change rate of VMaSC is lower than that of NHop and MDMAC in all cases, which again proves the higher stability attained by VMaSC. VMaSC reduces the cluster head change rate by leaving $CH$ state only when there is no member to serve whereas NHop and MDMAC use periodic clustering maintenance, which causes unnecessary cluster head change in the network. Moreover, VMaSC avoids unnecessary state transitions from $CH$ to $CM$ by ensuring the connectivity of two clusters for $MERGE\_TIMER$ before merging them. Furthermore, similar to cluster head and member duration, increasing the number of hops allowed in the clusters decreases cluster head change rate thus increases the stability of VMaSC. The cdf of the cluster head change rate on the other hand shows that the variation around the average cluster head change rate of VMaSC is again small demonstrating its stability.

*4) Clustering Overhead:* Clustering overhead is defined as the ratio of the total number of clustering related packets to the total number of packets generated in the VANET.

Fig. 7 shows the clustering overhead of the algorithms as a function of maximum vehicle velocities for different maximum number of hops. The clustering overhead of VMaSC is smaller than that of NHop and MDMAC. The first reason is better cluster stability of VMaSC with higher cluster head and member duration. Another reason is the efficient mechanism for connection to the cluster through the neighboring cluster member instead of connecting to the CH in multiple hops. The VMaSC also eliminates the overhead of periodic active clustering by a timer based cluster maintenance. Moreover, as the maximum velocity of the vehicles increases, the increase in the clustering overhead of NHop and MDMAC is steeper than that of VMaSC, which illustrates the stability of VMaSC in highly dynamic networks. Furthermore, the clustering overhead of the protocols increases as the maximum number of hops increases since the $HELLO\_PACKET$s are rebroadcast over multiple hops to the neighbors within $MAX\_HOP$.

*5) Number of Vehicles in SE state:* Fig. 8 shows the number of vehicles in $SE$ state of VMaSC as a function of simulation time for different maximum number of hops and different maximum vehicle velocities. The number of nodes in $SE$ state is larger at higher vehicle velocities when the maximum number of hops is small. This is expected since the connections in the network break with higher probability as the relative vehicle velocities increase. The difference between the number of nodes in $SE$ state at higher and lower vehicle velocities and the variation in the number of nodes in $SE$ state over time however decrease as the number of hops increases. The reasons for this decrease are 1) the suitability of a larger set of neighboring vehicles at higher maximum number of hops allowing better cluster selection for CMs and 2) the suitability of a larger number of cluster members decreasing the probability of losing all members and transitioning to $SE$ state for CHs.

### B. VANET-LTE Hybrid Architectures

The performance of the proposed VANET-LTE hybrid architecture, namely VMaSC-LTE, is compared to that of flooding; pure VANET cluster based data forwarding mechanisms including VMaSC, NHop and MDMAC where the cluster heads relay information over IEEE 802.11p network instead of eNodeBs; hybrid architectures NHop-LTE and MDMAC-LTE that integrate the VANET clustering algorithms NHop and MDMAC with LTE; and a recently proposed hybrid architecture named CMGM-LTE. CMGM-LTE is the adaptation of the clustering-based multi-metric adaptive gateway management mechanism (CMGM) proposed for UMTS [32] to LTE. CMGM-LTE uses clustering metric defined as a function of the received signal strength from base stations, direction of movement and inter-vehicular distance, and periodic cluster update based maintenance mechanism without any cluster merging.

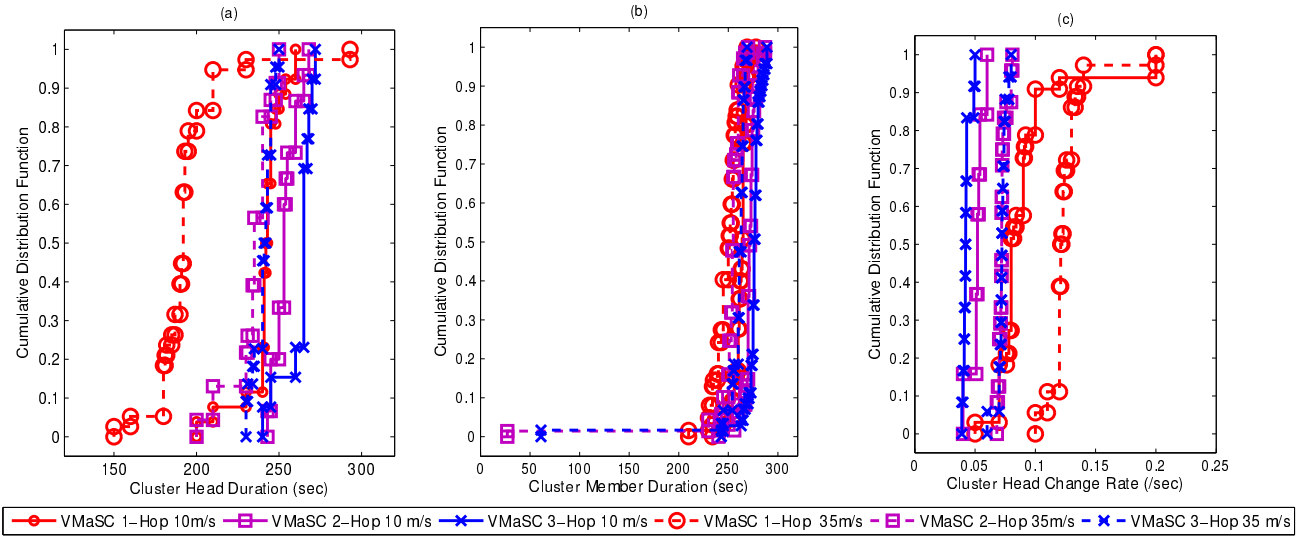

Fig. 6: CDF of (a) cluster head duration, (b) cluster member duration (c) cluster head change rate of VMaSC for different maximum number of hops and different maximum vehicle velocities.

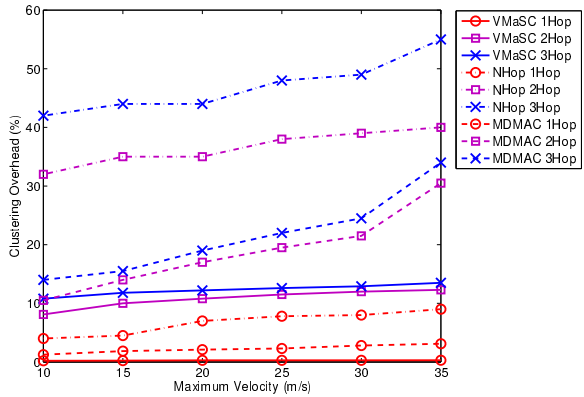

Fig. 7: Clustering overhead of the clustering algorithms for different maximum number of hops as a function of maximum vehicle velocities.

The performance metrics are data packet delivery ratio, delay and the cost of using LTE infrastructure.

*1) Data Packet Delivery Ratio (DPDR):* This metric is defined as the ratio of the number of the vehicles successfully receiving data packets to the total number of the vehicles within the target geographical area for the dissemination of the data packet. The average is taken over all the data packets sent by the vehicles in the simulation.

Fig. 9 shows the DPDR of different algorithms at different maximum velocities for 1, 2 and 3 hop based clustering mechanisms. The DPDR of VMaSC-LTE is above all the other algorithms in all cases. The reason for the superior DPDR performance of VMaSC-LTE over the other hybrid architectures, namely CMGM-LTE and MDMAC-LTE, is better clustering stability, minimal clustering overhead and minimal overlap among clusters. Higher clustering stability results in stable connections among cluster members and smaller number of nodes in $SE$ state. Minimal clustering overhead and minimal overlap among clusters on the other hand decrease the medium access contention increasing the success probability of the transmissions. The reason for the superior performance of the clustering based hybrid architectures over pure clustering based data forwarding and pure flooding on the other hand is the efficiency of LTE based communication among clusters in hybrid architectures. Hybrid architecture decreases the number of transmissions in IEEE 802.11p based network by providing LTE based inter-cluster communication, which in turn decreases the medium access contention. Moreover, the DPDR of VMaSC-LTE does not change considerably when the maximum velocity increases, demonstrating the robustness of the clustering algorithm to the increasing dynamicity of the network. Furthermore, the DPDR of VMaSC-LTE improves slightly as the maximum number of hops increases due to again higher clustering stability as demonstrated in Section V-A.

Fig. 10 shows the DPDR of VMaSC and VMaSC-LTE at different vehicle densities. The performance of pure cluster based data forwarding mechanism VMaSC is poor at low and high vehicle densities due to the disconnected network and broadcast storm problems, respectively. We observe that LTE based hybrid architecture VMaSC-LTE improves the performance greatly providing high DPDR stable at all vehicle traffic densities.

*2) Delay:* The delay metric is defined as the average latency of the data packets that travel from their source to the vehicles within the target geographical area of dissemination. The average is taken over both the packets and the destinations. On the other hand, the maximum delay metric is defined as the maximum latency of the data packets that travel from their source to the vehicles within the target geographical area of dissemination.

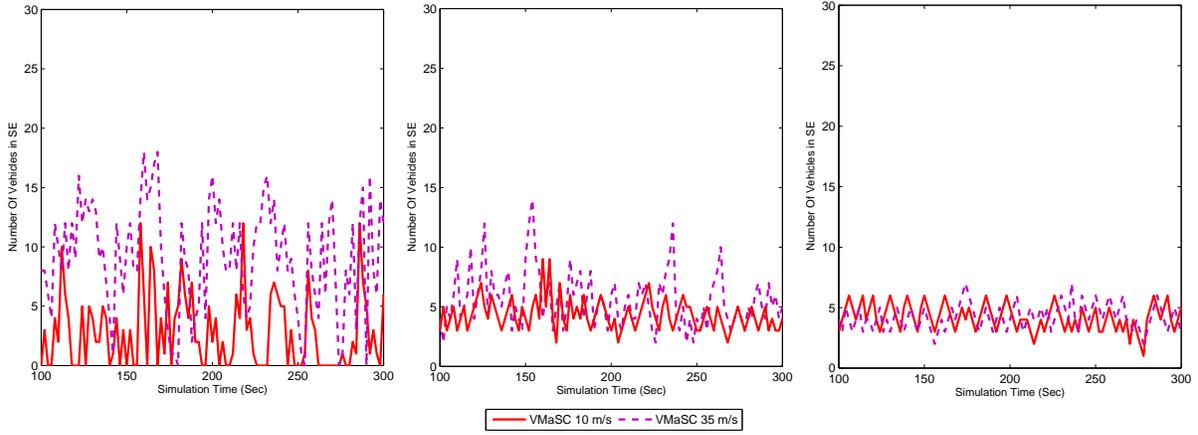

Fig. 8: Number of vehicles in $SE$ state for (a) 1-hop, (b) 2-hop (c) 3-hop VMaSC clustering

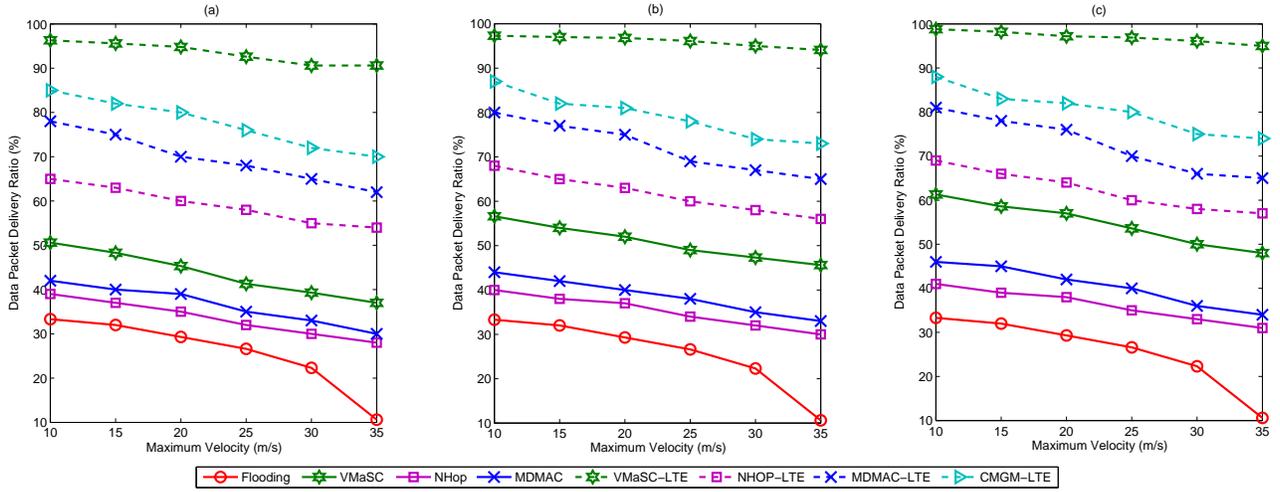

Fig. 9: DPDR of data dissemination algorithms at different maximum velocities for (a) 1-hop, (b) 2-hop (c) 3-hop based clustering

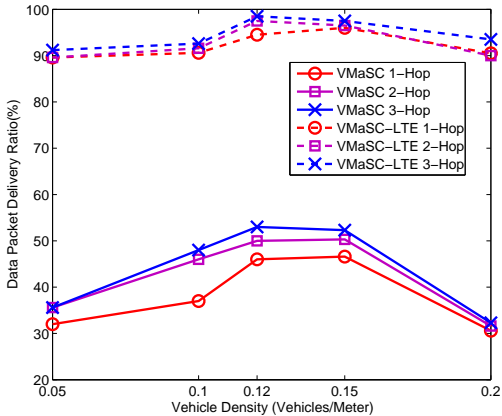

Fig. 10: DPDR of VMaSC and VMaSC-LTE at different vehicle densities.

Figs. 11 and 12 show the average and maximum delay of different algorithms at different maximum velocities for 1,2 and 3 hop based clustering mechanisms respectively. When we consider these results together with Fig. 9, we observe that there is a trade-off between DPDR and delay for flooding and pure cluster based algorithms: Flooding provides lower delay than cluster based algorithms whereas cluster based algorithms achieve higher DPDR than the flooding. LTE based hybrid architectures on the other hand achieve both low delay and high DPDR at the cost of using the infrastructure. Among the hybrid architectures, VMaSC-LTE achieves the lowest delay. Furthermore, the DPDR and delay analysis at different number of maximum hops allowed within clusters show that increasing the maximum number of hops increases the DPDR at the cost of slight increase in the delay.

*3) LTE Cost:* LTE cost metric indicates the cost of using LTE infrastructure to improve the data delivery performance of the hybrid architecture and is measured by the number of cluster heads in the network. The number of cluster heads depends on both the number of hops used in the clustering algorithm and the constraint on the maximum number of members CH and CM can admit, denoted by $MEMBER_{CH}$ and $MEMBER_{CM}$ respectively. We assume that the value of $MEMBER_{CH}$ varies from 1 to 10 however the value of $MEMBER_{CM}$ is kept constant at 1 in the simulation.

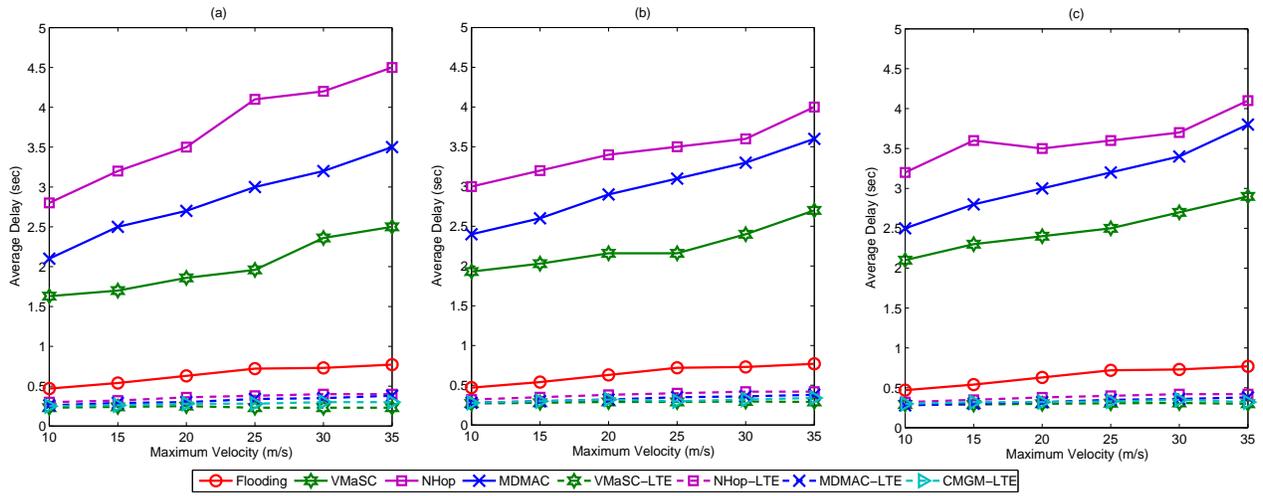

Fig. 11: Average delay of data dissemination algorithms at different maximum velocities for (a) 1-hop, (b) 2-hop (c) 3-hop based clustering.

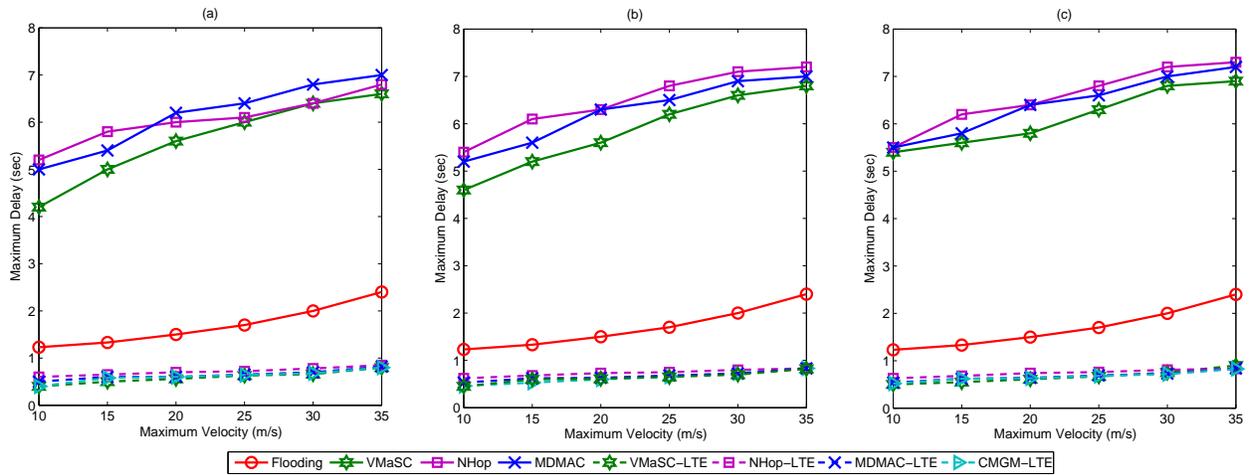

Fig. 12: Maximum delay of data dissemination algorithms at different maximum velocities for (a) 1-hop, (b) 2-hop (c) 3-hop based clustering.

TABLE VII: Number of clusters (LTE cost) of VMaSC-LTE for different $MEMBER_{CH}$ and number of hops values.

|  | Number of Clusters | | |
|---|---|---|---|
| $MEMBER_{CH}$ | 1-Hop | 2-Hop | 3-Hop |
| 1 | 72 | 65 | 48 |
| 3 | 52 | 50 | 49 |
| 5 | 42 | 38 | 35 |
| 10 | 32 | 42 | 38 |

Table VII shows the number of clusters, i.e. number of cluster heads, for different $MEMBER_{CH}$ and number of hops values. As $MEMBER_{CH}$ increases, the number of clusters in general decreases. The number of clusters however may increase back when the number of hops and $MEMBER_{CH}$ are high, e.g. $MEMBER_{CH} = 10$ at 2-hop scenario, due to the higher clustering overhead at the CHs, higher contention around CHs and larger number of CMs affected when a link within the cluster breaks. When both the number of hops and $MEMBER_{CH}$ are high in a cluster, the number of vehicles connected or sending a request for connection to the CH is higher. Moreover, the number of packets traveling to the CH is larger. Therefore, there exists a lot of contention around the CH. This results in the loss of packets around CH so loss of connections to the parent node in the routing path to the CH. When the connection of a vehicle to its parent node breaks, the vehicle and all of its children transition back to $SE$ state, increasing clustering overhead further and possibly creating unnecessarily higher number of clusters.

Fig. 13 shows the dependence of the DPDR on the $MEMBER_{CH}$ for different number of maximum allowed hops. We observe that the DPDR increases up to 100 as $MEMBER_{CH}$ decreases. The main reason for this behavior is the decrease in the clustering overhead and contention in the IEEE 802.11p based network with the decrease in $MEMBER_{CH}$. This demonstrates the adaptive usage of the VMaSC-LTE architecture depending on the reliability requirement of the application. As the reliability requirement of the application increases, the value of the $MEMBER_{CH}$ parameter needs to decrease at the cost of creating a larger number of clusters so increasing the cost of LTE usage.

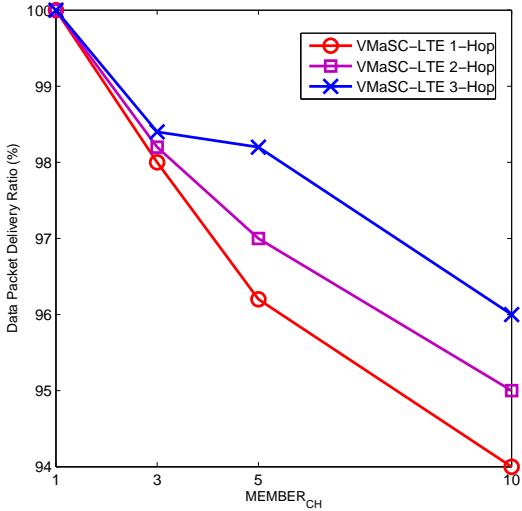

Fig. 13: DPDR of VMaSC-LTE for different $MEMBER_{CH}$ and number of hops values.

## VI. CONCLUSION

In this paper, we introduce a novel architecture VMaSC-LTE that integrates 3GPP/LTE networks with IEEE 802.11p based VANET networks. In VMaSC-LTE, vehicles are clustered in a multi-hop based novel approach named VMaSC with the features of cluster head selection using the relative mobility metric calculated as the average relative speed with respect to the neighboring vehicles, cluster connection with minimum overhead by introducing direct connection to the neighbor that is already a head or member of a cluster instead of connecting to the cluster head in multiple hops, disseminating cluster member information within periodic hello packets, reactive clustering to maintain cluster structure without excessive consumption of network resources, and efficient size and hop limited cluster merging mechanism based on the exchange of the cluster information among the cluster heads. In the constructed clusters, cluster heads activate the LTE interface to connect the VANET network to LTE.

Extensive simulations in ns-3 with the vehicle mobility input from SUMO demonstrate the superior performance of VMaSC-LTE over both previously proposed hybrid architectures and alternative routing mechanisms including flooding and cluster based routing. We observe that the DPDR performance of pure cluster based data forwarding mechanism is poor at low and high vehicle densities due to the disconnected network and broadcast storm problems respectively. LTE based hybrid architecture however improves the performance greatly providing high DPDR stable at all vehicle traffic densities. Moreover, despite the trade-off between DPDR and delay observed for flooding and pure cluster based algorithms, the proposed architecture has been demonstrated to achieve both low delay and high DPDR at the cost of using the LTE infrastructure. Among the hybrid architectures, VMaSC-LTE achieves the lowest delay and highest DPDR due to better clustering stability, minimal clustering overhead and minimal overlap among clusters. The DPDR and delay analysis at different number of maximum hops allowed within clusters shows that increasing the maximum number of hops up to 3 increases the DPDR at the cost of slight increase in the delay.

We have also defined LTE cost metric as the cost of using LTE infrastructure to improve the data delivery performance of the hybrid architecture. The LTE cost is measured by the number of cluster heads in the network. We observe that the DPDR increases up to 100 as the number of members allowed in the clusters decreases. The main reason for this behavior is the decrease in the clustering overhead and contention in the IEEE 802.11p based network with the decrease in the number of cluster members. This demonstrates the adaptive usage of the VMaSC-LTE architecture depending on the reliability requirement of the application. As the required reliability of the application increases, the number of cluster members needs to decrease at the cost of creating a larger number of clusters so increasing the cost of LTE usage.

As future work, we aim to investigate the use of VMaSC-LTE in urban traffic scenarios and extend VMaSC-LTE architecture with data aggregation and calculation of the clustering metric with additional information such as the most probable path information of the vehicles.

## REFERENCES

[1] S. Ucar, S. C. Ergen, and O. Ozkasap, "VMaSC: Vehicular multi-hop algorithm for stable clustering in vehicular ad hoc networks," in *Wireless Communications and Networking Conference (WCNC), IEEE*, 2013, pp. 2381–2386.
[2] R. Chen, W.-L. Jin, and A. Regan, "Broadcasting safety information in vehicular networks: issues and approaches," *Network, IEEE*, vol. 24, no. 1, pp. 20 –25, 2010.
[3] "Vehicle Safety Communications Project Task 3 Final Report: Identify Intelligent Vehicle Safety Applications Enabled byDSRC," in *The CAMP Vehicle Safety Communications Consortium,*, June 2010.
[4] "Intelligent Transport Systems (ITS); Vehicular Communications; Basic Set of Applications; Definitions," in *The CAMP Vehicle Safety Communications Consortium,*, June 2010.
[5] N. Wisitpongphan, O. Tonguz, J. Parikh, P. Mudalige, F. Bai, and V. Sadekar, "Broadcast storm mitigation techniques in vehicular ad hoc networks," *Wireless Communications, IEEE*, vol. 14, no. 6, pp. 84 –94, December 2007.
[6] N. Wisitpongphan, F. Bai, P. Mudalige, V. Sadekar, and O. Tonguz, "Routing in sparse vehicular ad hoc wireless networks," *Selected Areas in Communications, IEEE Journal on*, vol. 25, no. 8, pp. 1538–1556, 2007.
[7] T. Song, W. Xia, T. Song, and L. Shen, "A cluster-based directional routing protocol in VANET," in *Communication Technology (ICCT), 12th IEEE International Conference on*, 2010, pp. 1172–1175.
[8] B. Wiegel, Y. Gunter, and H. Grossmann, "Cross-layer design for packet routing in vehicular ad hoc networks," in *Vehicular Technology Conference VTC Fall. IEEE 66th*, 2007, pp. 2169–2173.
[9] Z. Zhang, A. Boukerche, and R. Pazzi, "A novel multi-hop clustering scheme for vehicular ad-hoc networks," in *Proceedings of the 9th ACM international symposium on Mobility management and wireless access*, 2011, pp. 19–26.
[10] N. Maslekar, M. Boussedjra, J. Mouzna, and L. Houda, "Direction based clustering algorithm for data dissemination in vehicular networks," in *Vehicular Networking Conference (VNC), IEEE*, 2009, pp. 1–6.
[11] H. Su and X. Zhang, "Clustering-based multichannel MAC protocols for QoS provisionings over vehicular ad hoc networks," *Vehicular Technology, IEEE Transactions on*, vol. 56, no. 6, pp. 3309–3323, 2007.
[12] M. Venkata, M. Pai, R. Pai, and J. Mouzna, "Traffic monitoring and routing in VANETs- a cluster based approach," in *ITS Telecommunications (ITST), 2011 11th International Conference on*, 2011, pp. 27–32.

[13] Y. Zhang and J. M. Ng, "A distributed group mobility adaptive clustering algorithm for mobile ad hoc networks," in *Communications. ICC. IEEE International Conference on*, 2008, pp. 3161–3165.
[14] Z. Y. Rawashdeh and S. Mahmud, "A novel algorithm to form stable clusters in vehicular ad hoc networks on highways." in *EURASIP J. Wireless Comm. and Networking*, 2012, p. 15.
[15] A. Daeinabi, A. G. Pour Rahbar, and A. Khademzadeh, "VWCA: an efficient clustering algorithm in vehicular ad hoc networks," *J. Netw. Comput. Appl.*, vol. 34, no. 1, pp. 207–222, Jan. 2011.
[16] G. Wolny, "Modified DMAC clustering algorithm for VANETs," in *Systems and Networks Communications. ICSNC*, 2008, pp. 268–273.
[17] Z. Wang, L. Liu, M. Zhou, and N. Ansari, "A Position-Based Clustering Technique for Ad Hoc Intervehicle Communication," *Systems, Man, and Cybernetics, Part C: Applications and Reviews, IEEE Transactions on*, March 2008.
[18] B. Hassanabadi, C. Shea, L. Zhang, and S. Valaee, "Clustering in Vehicular Ad Hoc Networks using Affinity Propagation," *Ad Hoc Networks*, vol. 13, Part B, no. 0, pp. 535 – 548, 2014.
[19] E. Souza, I. Nikolaidis, and P. Gburzynski, "A New Aggregate Local Mobility ALM; Clustering Algorithm for VANETs," in *Communications (ICC), IEEE International Conference on*, May 2010, pp. 1–5.
[20] O. Tonguz, N. Wisitpongphan, F. Bai, P. Mudalige, and V. Sadekar, "Broadcasting in VANET," in *Mobile Networking for Vehicular Environments*, 2007, pp. 7–12.
[21] O. Tonguz, N. Wisitpongphan, and F. Bai, "DV-CAST: A distributed vehicular broadcast protocol for vehicular ad hoc networks," *Wireless Communications, IEEE*, vol. 17, no. 2, pp. 47–57, April 2010.
[22] Project Cooperative Cars, CoCar @ONLINE. [Online]. Available: http://www.aktiv-online.org/english/aktiv-cocar.html/
[23] LTE-Connected Cars, ng Connect Program @ONLINE. [Online]. Available: http://www.ngconnect.org/index.htm/
[24] H. Abid, T.-C. Chung, S. Lee, and S. Qaisar, "Performance analysis of LTE smartphones-based vehicle-to-infrastrcuture communication," in *Ubiquitous Intelligence Computing and 9th International Conference on Autonomic Trusted Computing (UIC/ATC), 9th International Conference on*, 2012, pp. 72–78.
[25] G. Araniti, C. Campolo, M. Condoluci, A. Iera, and A. Molinaro, "LTE for vehicular networking: a survey," *Communications Magazine, IEEE*, vol. 51, no. 5, pp. 148–157, May 2013.
[26] I. Lequerica, P. Ruiz, and V. Cabrera, "Improvement of vehicular communications by using 3G capabilities to disseminate control information," *Network, IEEE*, vol. 24, no. 1, pp. 32–38, 2010.
[27] Traffic and Network Simulation Environment, TranNS @ONLINE. [Online]. Available: http://trans.epfl.ch/
[28] Simulation of Urban MObility, SUMO @ONLINE. [Online]. Available: http://http://sumo.sourceforge.net/
[29] G. Remy, S. M. Senouci, F. Jan, and Y. Gourhant, "LTE4V2X: LTE for a centralized VANET organization," in *Global Telecommunications Conference, GLOBECOM IEEE*, 2011, pp. 1–6.
[30] M. Fiore, J. Harri, F. Filali, and C. Bonnet, "Vehicular mobility simulation for VANETs," in *Simulation Symposium, ANSS. 40th Annual*, 2007, pp. 301–309.
[31] T. Taleb and A. Benslimane, "Design guidelines for a network architecture integrating VANET with 3G & beyond networks," in *Global Telecommunications Conference, GLOBECOM IEEE*, 2010, pp. 1–5.
[32] A. Benslimane, T. Taleb, and R. Sivaraj, "Dynamic clustering-based adaptive mobile gateway management in integrated VANET - 3G heterogeneous wireless networks," *Selected Areas in Communications, IEEE Journal on*, vol. 29, no. 3, pp. 559–570, 2011.
[33] R. Sivaraj, A. Gopalakrishna, M. Chandra, and P. Balamuralidhar, "QoS-enabled group communication in integrated VANET - LTE heterogeneous wireless networks," in *Wireless and Mobile Computing, Networking and Communications (WiMob), IEEE 7th International Conference on*, 2011, pp. 17–24.
[34] A. Benslimane, S. Barghi, and C. Assi, "An efficient routing protocol for connecting vehicular networks to the internet," *Pervasive and Mobile Computing*, pp. 98–113, 2011.
[35] F. Karnadi, Z. H. Mo, and K. chan Lan, "Rapid Generation of Realistic Mobility Models for VANET," in *IEEE Wireless Communications and Networking Conference (WCNC)*, March 2007.
[36] J. Ng and Y. Zhang, "A mobility model with group partitioning for wireless ad hoc networks," in *Information Technology and Applications, ICITA. Third International Conference on*, vol. 2, 2005, pp. 289–294.
[37] R. S. Bali, N. Kumar, and J. J. Rodrigues, "Clustering in vehicular ad hoc networks: Taxonomy, challenges and solutions," *Vehicular Communications*, vol. 1, no. 3, pp. 134 – 152, 2014.
[38] J. Yu and P. Chong, "3hBAC (3-hop between adjacent clusterheads): a novel non-overlapping clustering algorithm for mobile ad hoc networks," in *Communications, Computers and signal Processing. PACRIM. 2003 IEEE Pacific Rim Conference on*, vol. 1, pp. 318 – 321 vol.1.
[39] A. Damnjanovic, J. Montojo, Y. Wei, T. Ji, T. Luo, M. Vajapeyam, T. Yoo, O. Song, and D. Malladi, "A survey on 3GPP heterogeneous networks," *Wireless Communications, IEEE*, vol. 18, no. 3, pp. 10–21, June 2011.
[40] S. Ucar, S. C. Ergen, and O. Ozkasap, "VeSCA: Vehicular Stable Cluster-based Data Aggregation," in *International Conference on Connected Vehicles and Expo (ICCVE)*, November 2014.
[41] S. Yousefi, E. Altman, R. El-Azouzi, and M. Fathy, "Analytical Model for Connectivity in Vehicular Ad Hoc Networks," *Vehicular Technology, IEEE Transactions on*, vol. 57, no. 6, pp. 3341–3356, Nov 2008.
[42] M. Rudack, M. Meincke, and M. Lott, "On the Dynamics of Ad-Hoc Networks for Inter Vehicle Communications," in *International Conference on Wireless Networks (ICWN)*, 2002.
[43] R. S. Rao, S. K. Soni, N. Singh, and O. Kaiwartya, "A Probabilistic Analysis of Path Duration Using Routing Protocol in VANETs," *International Journal of Vehicular Technology*, 2014.
[44] F. Bai, N. Saragopan, and A. Helmy, "The important framework for analyzing the Impact of Mobility on Performance Of RouTing protocols for Adhoc NeTworks," *Ad Hoc Networks*, vol. 1, pp. 383–403, 2003.
[45] S. Al-Sultan, M. M. Al-Doori, A. H. Al-Bayatti, and H. Zedan, "A Comprehensive Survey on Vehicular Ad Hoc Network," *Journal of Network and Computer Applications*, vol. 37, pp. 380–392, January 2014.
[46] Network simulator, ns-3 @ONLINE. [Online]. Available: http://www.nsnam.org/
[47] M. Lacage and T. R. Henderson, "Yet another network simulator," in *Proceeding from the Workshop on Ns-2: The IP Network Simulator*. New York, NY, USA: ACM, 2006.